\documentclass[sigconf,screen]{acmart}
\makeatletter
\providecommand{\bibliofont}{}
\renewcommand{\bibliofont}{\scriptsize}
\makeatother
\usepackage{booktabs}
\usepackage{graphicx}
\usepackage{xcolor}
\usepackage{balance}

\setcopyright{cc}
\setcctype{by}
\acmDOI{10.1145/3832783.3834602}
\acmYear{2026}
\copyrightyear{2026}
\acmISBN{979-8-4007-2882-2/2026/10}
\acmConference[ASE '26]{Proceedings of the 41st IEEE/ACM International Conference on Automated Software Engineering}{October 12--16, 2026}{Munich, Germany}
\acmBooktitle{Proceedings of the 41st IEEE/ACM International Conference on Automated Software Engineering (ASE '26), October 12--16, 2026, Munich, Germany}
\acmSubmissionID{ase26tool-p28-p}
\received{2026-05-09}
\received[accepted]{2026-06-19}

\begin{document}

\title{FairLint-DL: An IDE-Native Tool for Fairness Debugging of Deep Learning Software}

\author{Archit Rathod}
\orcid{0009-0004-5498-1163}
\affiliation{%
  \institution{University of Illinois at Chicago}
  \city{Chicago}
  \country{USA}
}
\email{arath21@uic.edu}

\author{Saeid Tizpaz-Niari}
\orcid{0000-0002-1375-3154}
\affiliation{%
  \institution{University of Illinois at Chicago}
  \city{Chicago}
  \country{USA}
}
\email{saeid@uic.edu}



\begin{abstract}
  Existing fairness analysis tools predominantly operate as post-training evaluation frameworks, requiring practitioners to complete the full model development lifecycle before assessing bias. We present \textsc{FairLint-DL}, a Visual Studio Code extension that implements a \emph{shift-left} approach to fairness testing by enabling pre-training, IDE-native bias detection directly on tabular datasets. \textsc{FairLint-DL} trains a configurable deep neural network as a proxy model and applies information-theoretic Quantitative Individual Discrimination (QID) metrics. Grounded in Shannon and min-entropy, QID quantifies the causal influence of protected attributes on predictions. The system implements a two-phase gradient-guided search algorithm for discovering discriminatory instances, a causal debugging pipeline that localizes bias to specific network layers and neurons via sensitivity analysis, and dual explainability engines using SHAP and LIME for feature-level attribution. Evaluation on three tabular benchmarks (Adult Census Income, German Credit, and Bank Marketing) reveals fairness concerns that vary widely across datasets: on Adult, 96.0\% of analyzed instances exhibit QID above the 0.1-bit significance threshold, with a mean QID of 0.619 bits and a disparate impact ratio of 0.581, violating the four-fifths legal rule. \textsc{FairLint-DL} produces these results within 12 seconds on cached models, demonstrating the feasibility of integrating fairness analysis into the developer workflow without significant overhead.

  \smallskip
  \noindent\textbf{Demo video:} \url{https://youtu.be/AmfpjK24uwY} \newline \textbf{Source:} \url{https://github.com/Archit1706/FairLint-DL} \newline \textbf{Tool:} \url{https://open-vsx.org/extension/ArchitRathod/fairlint-dl}
\end{abstract}

\maketitle

\section{Introduction}

The widespread adoption of deep neural networks (DNNs) in high-stakes decision-making systems has introduced significant concerns regarding algorithmic fairness. Instances of such AI-assisted software include systems deciding on recidivism risk, software predicting benefit eligibility, and tools deciding whether to audit a given taxpayer~\cite{dice}. Such models often encode and amplify pre-existing biases in the training data. The resulting fairness defects may disadvantage protected groups and may also violate statutory requirements such as the Civil Rights Act's disparate impact provisions and the EU AI Act's fairness mandates.

\noindent \textbf{The Shift-Left Fairness Testing Paradigm.} Current fairness tools predominantly operate in a post-training paradigm: practitioners must complete the full model development lifecycle. Tools such as IBM's AI Fairness 360, Google's What-If Tool, Microsoft's Fairlearn, and Fairkit-learn~\cite{bellamy2019,Wexler_2019,fairlearn,fairkitLearn} provide valuable post-hoc analysis but require significant upfront investment before fairness issues are identified. Pre-training fairness assessment on datasets is itself well studied; our goal is not to claim it as new, but to make it a low-friction step inside the developer's editor. \textsc{FairLint-DL} applies this shift-left view to fairness testing, analogous to shift-left testing in software engineering. Delivered as a VS Code extension, it lets practitioners assess dataset bias before committing to a specific model architecture or training pipeline. The proxy DNN is trained once and cached as a fairness oracle for the dataset, so iteration cost on the developer's own model is zero and dataset-level bias signals (QID, group disparities) surface independently of the final architecture. Bias detected early is cheaper to fix; the feedback loop is tighter because results appear where code is written.

We use \emph{fairness debugging} to mean the workflow of detecting, quantifying, localizing, and explaining bias so that a developer can act on it, and we name the tool \textsc{FairLint-DL} because it brings this workflow to the point of editing, in the spirit of a linter, rather than performing static analysis on source code.

\noindent \textbf{Contributions.} The techniques we build on, namely QID, gradient-guided search, layer and neuron causal localization, and SHAP/LIME attribution, are established in prior work~\cite{dice,neufair,lundberg2017,ribeiro2016}; the contribution of this paper is their integration into a single editor-native, pre-training workflow rather than the techniques themselves. Concretely, we contribute: (1) an editor-integrated tool that packages pre-training fairness debugging behind a right-click on a dataset; (2) an end-to-end pipeline combining information-theoretic QID metrics, a two-phase discriminatory-instance search, layer and neuron localization, dual SHAP/LIME attribution, and a composite fairness score; and (3) an evaluation on three tabular fairness benchmarks.

\section{Background}

\textbf{Fairness Definitions.} Individual fairness, formalized by Dwork et al.~\cite{dwork2012}, requires that similar individuals receive similar outcomes. Group fairness requires parity of some statistical measure across demographic groups. \textsc{FairLint-DL} computes Demographic Parity, Equalized Odds~\cite{hardt2016}, and Equal Opportunity. These metrics are complementary: well-known impossibility results show that calibration, equalized odds, and unequal base rates cannot be jointly satisfied~\cite{kleinberg2016,chouldechova2017}, motivating the multi-metric approach.

\noindent \textbf{State-of-the-art Tools.} The DICE framework~\cite{dice} introduced Quantitative Individual Discrimination (QID) as an information-theoretic characterization of discrimination, and a causal debugging algorithm that uses interventions on neuron activations to localize layers and neurons with significant causal effects on QID. NeuFair~\cite{neufair} extends the search via a two-phase global-local approach with gradient-based optimization, and formulates fairness repair as simulated annealing over neuron dropout.
\textsc{FairLint-DL} also draws on the EVT framework~\cite{evt}, which applies extreme value theory to worst-case discrimination.

Table~\ref{tab:positioning} compares \textsc{FairLint-DL} with representative tools. AI Fairness 360 and Fairlearn are Python APIs that practitioners instrument after training; the What-If Tool is a post-training GUI; Fairkit-learn targets model selection on the fairness-accuracy frontier; FairLay-ML~\cite{fairlayml} is a separate web application for scikit-learn classifiers; and DICE~\cite{dice} and NeuFair~\cite{neufair}, the closest research artifacts, are CLI tools tied to benchmark scripts. \textsc{FairLint-DL} combines the information-theoretic rigor of DICE with the search of NeuFair behind an editor-native workflow.

\begin{table}[!tb]
  \centering
  \caption{Positioning of \textsc{FairLint-DL} relative to existing fairness tools. \emph{IDE}: editor-native integration; \emph{Pre}: pre-training focus; \emph{DNN}: DNN causal localization, i.e., attributing bias to specific layers and neurons of a deep network; \emph{QID}: information-theoretic QID; \emph{GS}: gradient-guided search; \emph{Grp}: group fairness; \emph{XAI}: SHAP/LIME. These dimensions describe the design space \textsc{FairLint-DL} targets, not an exhaustive ranking; toolkits such as AI Fairness 360 and Fairlearn expose many statistical fairness metrics that \textsc{FairLint-DL} does not.}
  \label{tab:positioning}
  \scriptsize
  \setlength{\tabcolsep}{3.5pt}
  \begin{tabular}{lccccccc}
    \toprule
    \textbf{Tool} & \textbf{IDE} & \textbf{Pre} & \textbf{DNN} & \textbf{QID} & \textbf{GS} & \textbf{Grp} & \textbf{XAI} \\
    \midrule
    AI Fairness 360~\cite{bellamy2019}        & --- & --- & --- & --- & --- & \checkmark & part. \\
    Fairlearn~\cite{fairlearn}                & --- & --- & --- & --- & --- & \checkmark & --- \\
    What-If Tool~\cite{Wexler_2019}           & --- & --- & --- & --- & --- & \checkmark & part. \\
    Fairkit-learn~\cite{fairkitLearn}         & --- & --- & --- & --- & --- & \checkmark & --- \\
    FairLay-ML~\cite{fairlayml}               & --- & --- & --- & --- & --- & \checkmark & LIME \\
    DICE (CLI)~\cite{dice}                    & --- & --- & \checkmark & \checkmark & --- & --- & --- \\
    NeuFair (CLI)~\cite{neufair}              & --- & --- & \checkmark & --- & \checkmark & \checkmark & --- \\
    \textbf{\textsc{FairLint-DL}}             & \checkmark & \checkmark & \checkmark & \checkmark & \checkmark & \checkmark & \checkmark \\
    \bottomrule
  \end{tabular}
\end{table}

\section{System Architecture and Design}

\begin{figure}[!tb]
  \vspace{2mm}
  \centering
  \includegraphics[width=\linewidth]{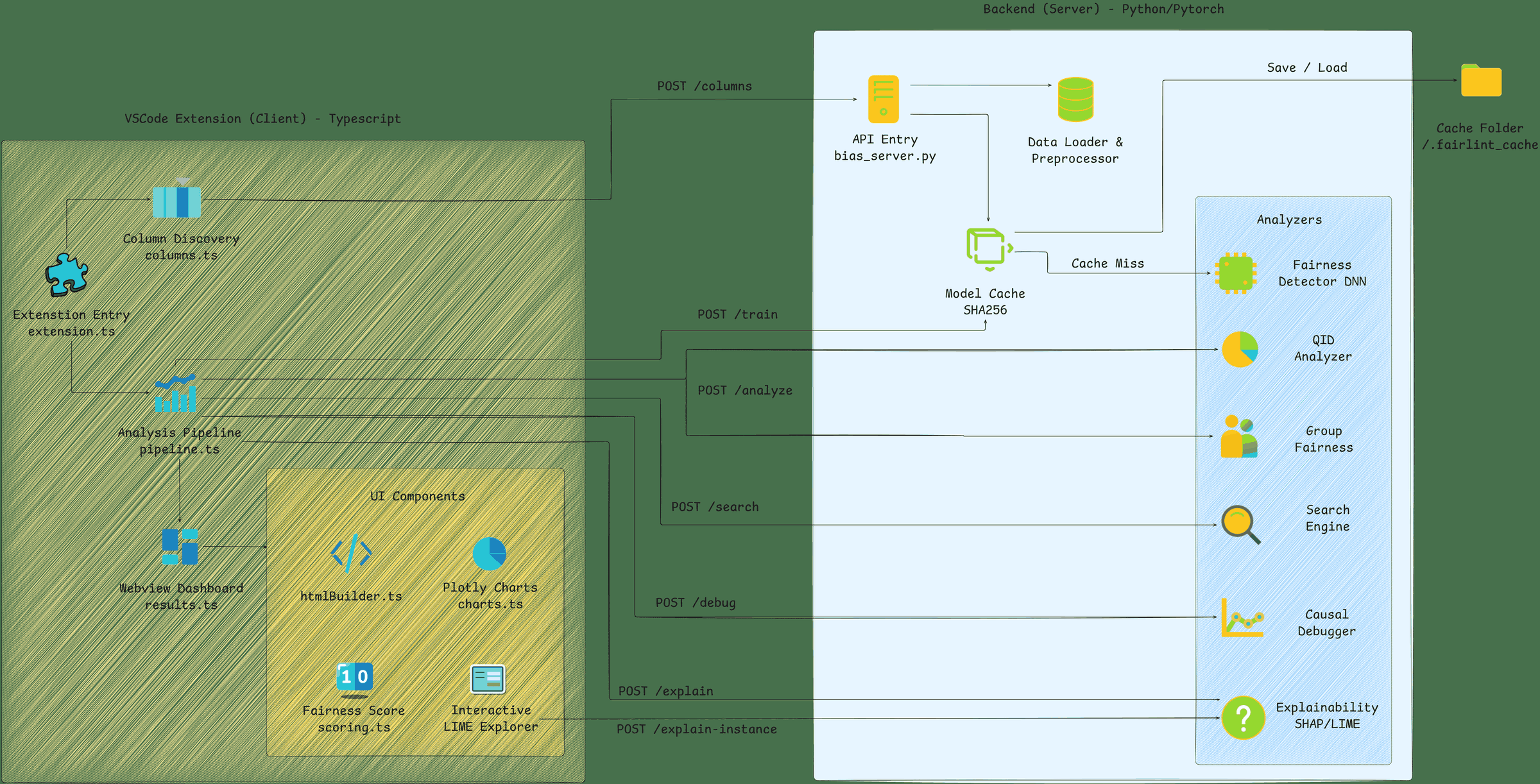}
  \caption{\textsc{FairLint-DL} client-server architecture overview.}
  \Description{System architecture diagram showing the VS Code extension on the client side communicating over localhost HTTP REST with a Python FastAPI backend that runs model training, QID analysis, discriminatory instance search, causal debugging, and explainability modules.}
  \label{fig:arch}
\end{figure}

\textsc{FairLint-DL} operates as a client-server application with two primary components communicating over HTTP/REST on \texttt{localhost:8765} (Figure~\ref{fig:arch}). The \textbf{VS Code Extension (Client)}, written in TypeScript, handles all user interaction, file system access, WebView dashboard rendering with Plotly.js charts, and manages the lifecycle of the Python backend. The \textbf{Python FastAPI Backend (Server)} hosts the PyTorch models and all analysis algorithms, including data preprocessing, model training, QID computation, causal debugging, discriminatory instance search, and SHAP/LIME explainability. The extension launches and manages the backend as a local process, so no separate server deployment is required. Because the two components communicate only over HTTP/REST, the tool is not bound to VS Code: the extension runs unmodified in any VS Code-compatible IDE, including Cursor and Antigravity, and the same endpoints can be driven by a CI job or a script.

\noindent \textbf{Six-Step Analysis Pipeline.} The pipeline is: model training (or cached load), PCA-based internal space visualization, QID and group fairness computation across all protected attributes~\cite{dice}, gradient-guided global plus perturbation-based local search for discriminatory instances~\cite{dice,neufair}, layer/neuron causal debugging~\cite{dice}, and batch SHAP (KernelExplainer on log-odds) plus LIME explanations~\cite{lundberg2017,ribeiro2016}.

\section{Algorithmic Methodology}

\textbf{DNN Proxy Model.} \textsc{FairLint-DL} trains a feedforward DNN as a proxy model to serve as a ``fairness oracle.'' The default architecture uses 5 fully-connected hidden layers of $\langle 64, 32, 16, 8, 4 \rangle$ neurons followed by a 2-unit softmax output head ($\sim$4K parameters total), following the architecture used in prior fairness testing literature~\cite{dice,neufair}. Each hidden layer applies Batch Normalization, ReLU activation, and Dropout ($p=0.2$). We train a DNN rather than run statistical tests on raw data because the differentiable model supports counterfactual and gradient-based search, learns non-linear feature interactions, and lets causal debugging trace bias through internal representations.

\noindent \textbf{QID Metrics.} For non-protected $X=x$ and protected $Z=z$, the Shannon entropy QID quantifies prediction uncertainty when varying $Z$ at fixed $X$:
\begin{equation}
  \mathrm{QID}_{\mathrm{Shannon}}(x) = -\sum_{y} \bar{P}(Y=y \mid x) \log_2 \bar{P}(Y=y \mid x)
\end{equation}
where $\bar{P}(Y=y \mid x)$ is the average prediction probability across all counterfactual variations of the protected attributes~\cite{dice}. \textsc{FairLint-DL} also computes the disparate impact ratio
\begin{equation}
  \mathrm{DI}(x) = \min_z P(\hat{Y}=1 \mid x,z) / \max_z P(\hat{Y}=1 \mid x,z)
\end{equation}

The four-fifths rule ($\mathrm{DI} \leq 0.8$) is the established U.S.\ civil rights threshold for adverse impact in employment discrimination; we keep it fixed by legal convention rather than as a tunable knob, whereas the QID significance threshold, epochs, batch size, layer sizes, and sample count are all user-configurable settings.

\noindent \textbf{Gradient-Guided Search.} The search for discriminatory instances is formulated as a two-phase optimization problem~\cite{dice,neufair}. Phase 1 (Global) maximizes the QID score directly using gradient ascent over input features. Phase 2 (Local) explores the neighborhood of the global optimum by adding Gaussian noise to non-protected features and filtering neighbors that exceed a threshold of $0.1$ bits.

\noindent \textbf{Causal Debugging.} Layer localization determines which layer is most sensitive to discriminatory inputs by computing the gradient of layer activations with respect to the input and aggregating the magnitude of these gradients for known discriminatory instances. Within the bias-critical layer, neuron localization follows the Average Causal Difference framework from DICE~\cite{dice}: the tool intervenes on each neuron by forcing it active or inactive and measures the resulting change in QID. Neurons with disproportionately high impact scores are candidates for pruning or repair.

\noindent \textbf{Composite Fairness Score.} A 0 to 100 score combines five penalty components (mean QID, \% discriminatory, disparate impact, max QID, group fairness). The weights are fixed constants in the current release and are listed in the artifact; exposing them as user settings is left to future work.
\section{Implementation and Usage}

\noindent \textbf{Technology Stack.} The backend uses PyTorch 2.0+ for model training and gradient computation, FastAPI + Uvicorn for the REST API, Pandas and Scikit-learn for data processing, and SHAP and LIME for feature attribution. The frontend is a TypeScript VS Code extension that renders interactive charts via Plotly.js.

\noindent \textbf{Model Caching.} To avoid redundant retraining, \textsc{FairLint-DL} caches trained models under a deterministic SHA-256 key computed from the CSV contents, label column, sorted sensitive features, hidden layer sizes, epochs, and batch size.

\begin{figure}[!tb]
  \centering
  \includegraphics[width=0.9\linewidth]{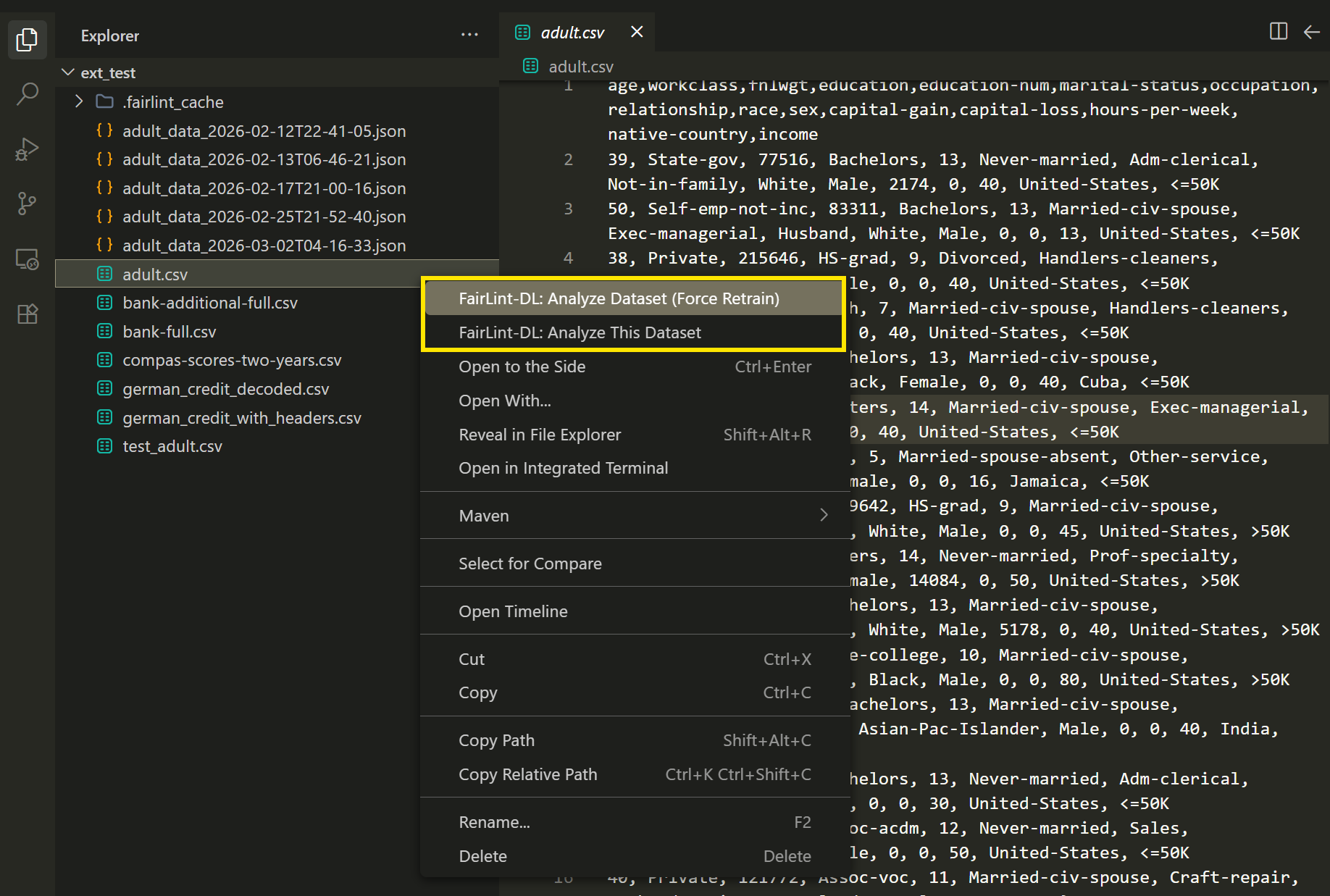}
  \caption{\textsc{FairLint-DL} integrates into the editor workflow.}
  \Description{Screenshot of Visual Studio Code with a CSV file selected in the explorer and a context menu option to launch FairLint-DL analysis, demonstrating how the extension is invoked from the editor.}
  \label{fig:ui}
\end{figure}

\noindent \textbf{Workflow.} Users trigger analysis via the context menu (right-click on a \texttt{.csv} file) (Figure~\ref{fig:ui}). The extension presents QuickPick dialogs for selecting the label column, protected attributes, and model architecture. Protected attributes are auto-detected by matching column names, case-insensitively, against a small curated set of regular expressions for common sensitive categories: sex or gender, race or ethnicity, age, nationality or country, religion, and disability. Auto-detection is only a default; the developer can deselect or add attributes in the QuickPick dialog. Configurable settings are exposed through VS Code's settings API, including training epochs, batch size, hidden layer configuration, QID threshold, maximum analysis samples, server port, and visualization theme.

\noindent \textbf{Availability.} \textsc{FairLint-DL} is MIT-licensed and published on the VS Code Marketplace and Open VSX for one-click install in any compatible IDE; from source, \texttt{npm install \&\& npm run compile} builds the extension and \texttt{pip install -r requirements.txt} the backend. The backend ships with a unit-test suite covering QID, search, causal debugging, and group fairness.

\section{Experimental Evaluation}

\textbf{Dataset.} The Adult Census Income dataset~\cite{adultdataset} contains 32{,}561 records from the 1994 U.S. Census Bureau database; the prediction task is binary classification of whether annual income exceeds \$50{,}000. Four protected attributes were selected: age (71 unique values), race (5), sex (2), and native-country (42). The dataset was split into 22{,}792 training, 3{,}256 validation, and 6{,}513 held-out test instances; the proxy DNN was trained for 30 epochs (training-set class distribution: 17{,}303 negative, 5{,}489 positive), and all fairness results below are computed on 500 instances sampled from the held-out test split. To test generalizability beyond a single benchmark, we apply the same architecture and protocol to German Credit~\cite{germancredit} (1{,}000 records; protected attributes age and sex) and Bank Marketing~\cite{bankmarketing} (45{,}211 records; protected attributes age and marital status). Table~\ref{tab:qid} reports QID and composite results for all three; the figures below detail the Adult case.

\begin{figure}[!tb]
  \centering
  \includegraphics[width=0.9\linewidth]{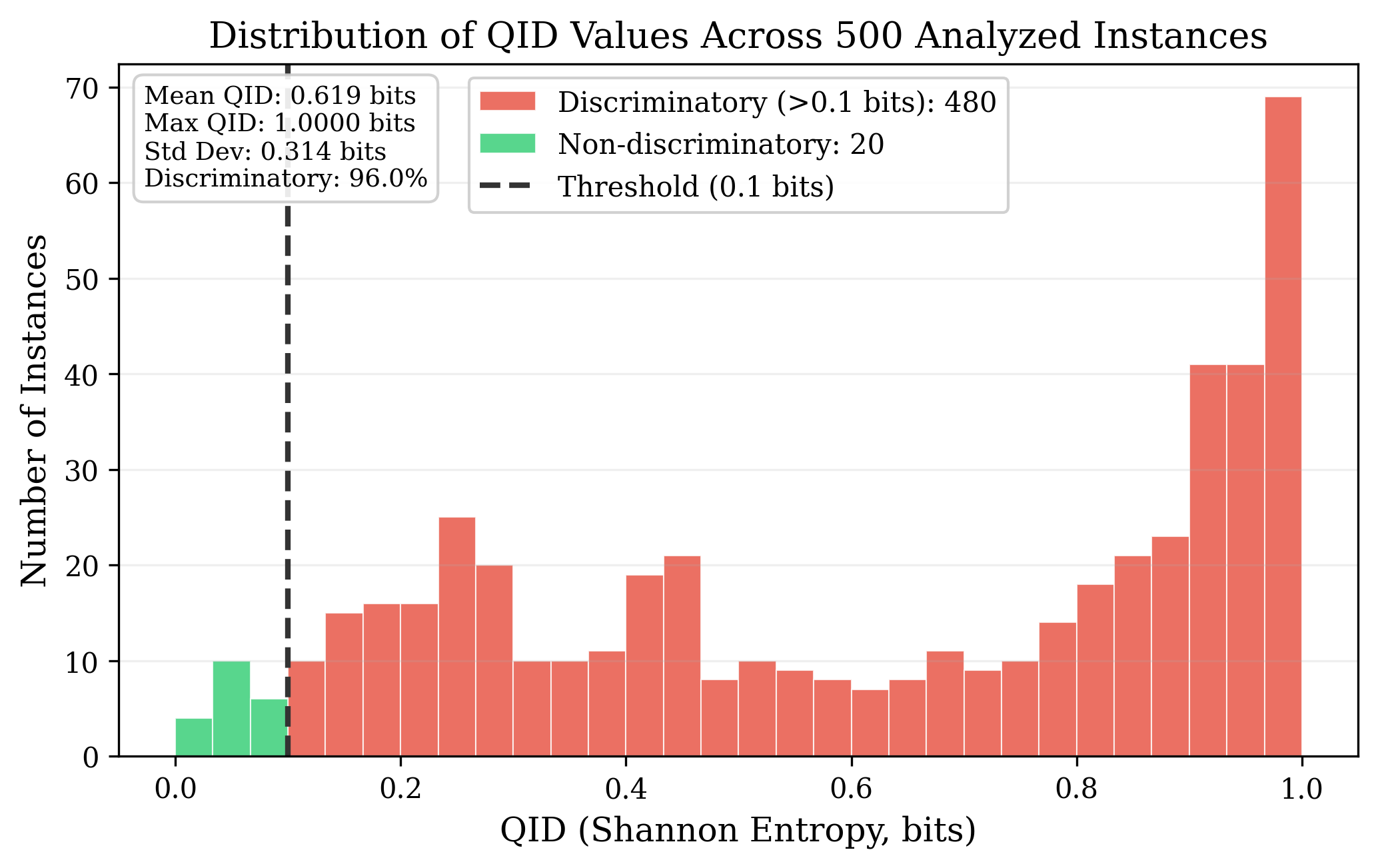}
  \caption{Distribution of QID values across 500 analyzed instances on Adult Census; 96.0\% exceed the 0.1-bit threshold.}
  \Description{Histogram of QID values on the Adult Census dataset, bimodal with a small cluster near zero and a large mass approaching 1.0 bit.}
  \label{fig:qid-dist}
\end{figure}

\noindent \textbf{QID Analysis.} On Adult, 500 sampled test instances reveal pervasive individual discrimination (Table~\ref{tab:qid}): a mean QID of 0.619 bits (of a 1.0-bit maximum) and a 96.0\% discrimination rate mean the model changes its prediction for nearly every individual when protected attributes are altered, and the mean disparate impact of 0.581 falls well below the 0.8 threshold. The QID histogram (Figure~\ref{fig:qid-dist}) is bimodal, with a small cluster near zero and a large mass approaching 1.0 bit, indicating that discrimination is a systemic property of the decision boundary rather than an edge-case phenomenon. The same pipeline surfaces markedly different profiles on the other benchmarks (Table~\ref{tab:qid}): German Credit is the most severe at the individual level (mean QID 0.935, all instances discriminatory), while Bank Marketing is the mildest (mean QID 0.284, 51.6\% discriminatory), giving composite scores of 46 and 75 against Adult's 41. German Credit also illustrates that individual and group views can disagree: nearly every instance is individually discriminatory even though its aggregate disparate impact ratio of 0.94 satisfies the four-fifths rule.

\begin{table}[!tb]
  \centering
  \caption{Individual (QID) and group fairness across three benchmarks (500 test instances each for Adult and Bank; 200 for German). Protected: Adult (age, race, sex, country), German (age, sex), Bank (age, marital). Group diffs are averaged over attributes; Bank's are near zero as its majority-class predictions give near-equal group rates.}
  \label{tab:qid}
  \small
  \setlength{\tabcolsep}{4pt}
  \begin{tabular}{lrrr}
    \toprule
    \textbf{Metric} & \textbf{Adult} & \textbf{German} & \textbf{Bank} \\
    \midrule
    Mean QID (Shannon, bits)        & 0.619  & 0.935   & 0.284 \\
    Mean min-entropy QID (bits)     & 0.214  & 0.428   & 0.108 \\
    Mean disparate impact ratio     & 0.581  & 0.939   & 0.815 \\
    Discriminatory (QID $>$ 0.1)    & 96.0\% & 100.0\% & 51.6\% \\
    Violating 80\% rule             & 69.8\% & 3.5\%   & 42.2\% \\
    Demographic parity diff.        & 0.091  & 0.095   & 0.000 \\
    Equalized odds diff.            & 0.102  & 0.114   & 0.000 \\
    Model accuracy (\%)             & 84.1   & 68.5    & 88.3 \\
    Composite fairness score        & 41     & 46      & 75 \\
    \bottomrule
  \end{tabular}
\end{table}

\noindent \textbf{Causal Debugging.} Layer 2 (32 neurons) is the most bias-sensitive layer (sensitivity 1.933), indicating that early-to-middle layers encode protected information before it propagates deeper. This matches the DICE results on Adult, which also identified the second layer~\cite{dice}. Within Layer 2, Neurons 30 (1.362), 17 (1.270), and 8 (1.220) have the highest impact and are prime candidates for the neuron-level dropout proposed by NeuFair~\cite{neufair}.


\begin{figure}[!tb]
  \centering
  \includegraphics[width=1.0\linewidth]{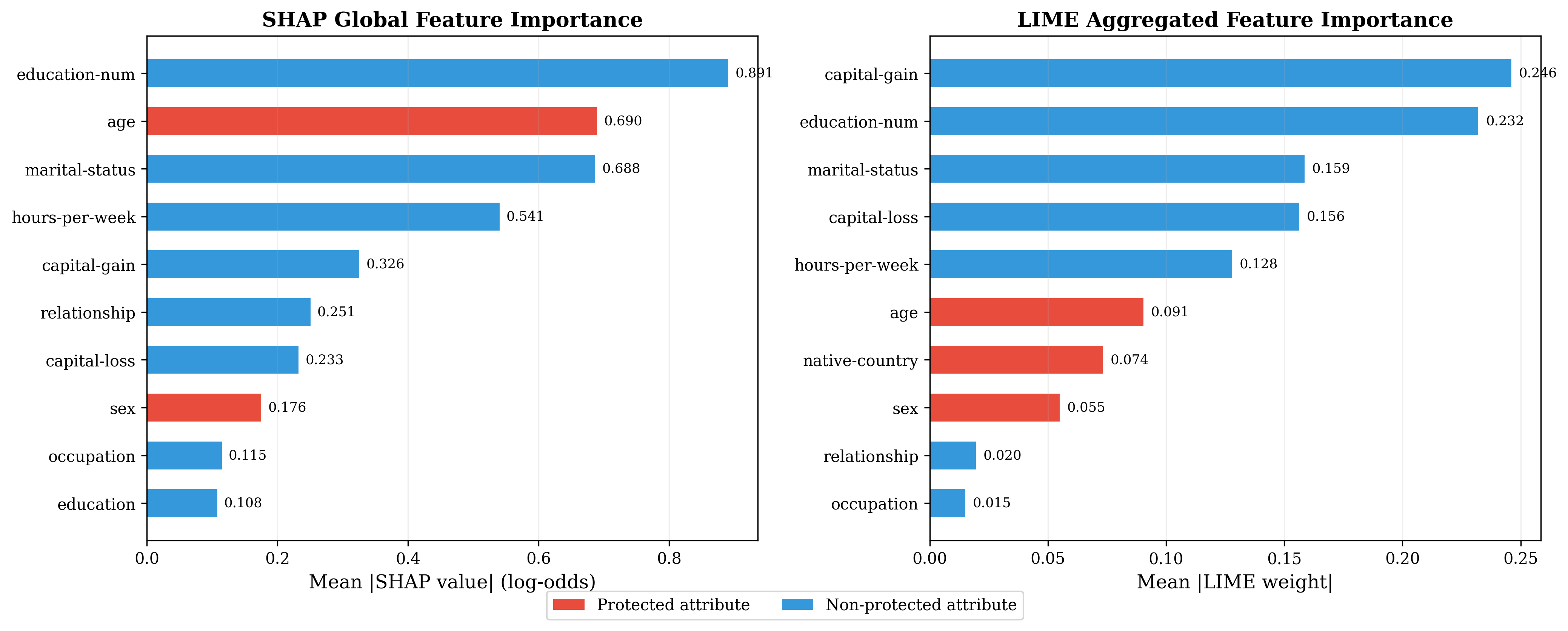}
  \caption{SHAP and LIME global feature importance on Adult Census; both methods agree that \texttt{education-num} and \texttt{age} (a protected attribute) are top drivers.}
  \Description{Bar charts comparing SHAP and LIME global feature importance on Adult Census, both ranking education-num and age among the top drivers.}
  \label{fig:shap-lime}
\end{figure}

\noindent \textbf{Explainability.} SHAP global importance (log-odds) ranks \texttt{education -num} (0.891), \texttt{age} (0.690), and \texttt{marital-status} (0.688) highest; \texttt{age}, a protected attribute, ranks second. SHAP and LIME agree on \texttt{education-num} and on the measurable influence of \texttt{age} (Figure~\ref{fig:shap-lime}), giving high confidence in the attribution.

\noindent \textbf{Composite Fairness Score and Performance.} Adult's final score of approximately 41 out of 100 is rated ``Concerning.'' With caching, the full analysis completes in about 12 (s) (SHAP+LIME dominates at $\sim$7s; other steps $\le$2s); first-run training adds about 77 (s).

\section{Discussion and Limitations}

\textbf{Interpretation.} On Adult, discrimination is a property of the learned decision boundary rather than an edge case. Although race shows minimal direct SHAP importance (0.010), the group fairness analysis reveals significant discrimination across racial groups (race-specific DP ratio 0.564), suggesting that race influences predictions indirectly through correlated proxy features, a form of proxy discrimination that single-method analysis may miss.

\textbf{Limitations.} \textsc{FairLint-DL} trains its own DNN proxy rather than analyzing a user-supplied model, so the detected bias reflects the proxy and may differ from a production model with a different architecture or fairness constraints; we have not yet compared QID on our proxy against alternative families such as gradient-boosted trees. The $\le 0$ vs.\ $> 0$ group split suits binary attributes but not continuous or multi-class ones, where quantile-based splits are a natural extension. SHAP's KernelExplainer scales quadratically in the number of features and is the bottleneck. Broader validation on intersectional attributes remains future work.

\section{Related Work}

\noindent\textbf{Fairness Testing.}
Counterfactual discrimination, introduced by Galhotra et al.~\cite{galhotra2017}, catalyzed a broad line of research on generating test cases at scale~\cite{SG-AggarwalESEC/FSE201910.1145/3338906.3338937,fanExpGAICSE22,BiswasICSE2023Fairify-10.1109/ICSE48619.2023.00134}.
One family of generators operates in a \emph{black-box} setting and requires no access to the internals of the classifier under test, including \textsc{AEQUITAS}~\cite{udeshi2018automated}, \textsc{SG}~\cite{SG-AggarwalESEC/FSE201910.1145/3338906.3338937}, \textsc{ExpGA}~\cite{fanExpGAICSE22}, \textsc{LIMI}~\cite{LIMI-XiaoISSTA2023-10.1145/3597926.3598099}, and \textsc{AFT}~\cite{AFT-zhao-ase2024}.
A second family adopts a \emph{white-box} setting and exploits gradient- or neuron-level information from the model, including \textsc{ADF}~\cite{zhang2020white}, \textsc{EIDIG}~\cite{EDIG-Zhang-ISSTA2021-10.1145/3460319.3464820}, \textsc{NeuronFair}~\cite{ZhengICSE2022NeuronFair}, \textsc{DICE}~\cite{dice}, and \textsc{MAFT}~\cite{MAFT-Wang-ICSE2024-10.1145/3597503.3639181}.
\textsc{FairLint-DL} builds on this white-box tradition, and its metrics are closely related to the information-theoretic view of discrimination~\cite{dice,11334575}. However, \textsc{FairLint-DL} surfaces IDIs early in development and directly within the IDE.
The complementary problem of group fairness testing has likewise been studied extensively~\cite{chakraborty2020fairway, DBLP:conf/icse/ZhangH21, 10.1145/3468264.3468537, tizpaz2022fairness-10.1145/3510003.3510202,10.1145/3663533.3664040}.

\noindent\textbf{Bias Mitigation.}
A substantial body of work mitigates bias in machine learning outputs through pre-, in-, and post-processing interventions, including post-processing methods that adjust model decisions after training~\cite{neufair,10.1145/3510003.3510080,10.1145/3617168,10.1145/3540250.3549103,10.1145/3510003.3510087,Dasu26Attn}.
These efforts are largely orthogonal and complementary to \textsc{FairLint-DL}.

\section{Conclusion and Future Work}

We presented \textsc{FairLint-DL}, a VS Code extension for deep learning-based fairness debugging of machine learning datasets. By implementing the shift-left fairness testing paradigm, \textsc{FairLint-DL} enables practitioners to detect, quantify, localize, and explain bias directly within their development environment. Future work includes integrating in-processing fairness interventions such as adversarial debiasing, fairness-constrained optimization~\cite{agarwal2018}, or the NeuFair dropout strategy~\cite{neufair} directly into the extension; extending evaluation to further datasets such as COMPAS; analyzing intersectional fairness across combinations of protected attributes; and supporting external model analysis beyond proxy models.

\section*{Acknowledgments}
This project has been partially supported by NSF award CCF-2536640.

\section*{Data Availability Statement}
The \textsc{FairLint-DL} source code, datasets, and evaluation scripts are archived on Zenodo~\cite{fairlintdl-artifact} (DOI: \url{https://doi.org/10.5281/zenodo.21362910}) under the MIT license; the archive includes \texttt{reproduce.py}, which regenerates Table~\ref{tab:qid} from the bundled datasets. The live source and packaged extension are linked in the abstract.

\balance

\bibliographystyle{ACM-Reference-Format}
\setlength{\bibsep}{0pt plus 0.2pt}
\bibliography{references}

\end{document}